\documentclass[aps,prb,twocolumn,superscriptaddress,showpacs,amsfonts]{revtex4-1}

\usepackage{epsfig}
\usepackage{graphicx}
\usepackage{amsmath,ulem}
\usepackage{amssymb}
\usepackage{color}
\usepackage{bm}

\begin{document}

\title{Proximity effects in spin-triplet superconductor-ferromagnet heterostucture with spin-active interface}

\author{Damien~Terrade}
\email[Electronic address: ]{d.terrade@fkf.mpg.de}
\affiliation{Max Planck Institute for Solid State Research,
D-70569 Stuttgart, Germany}
\author{Paola~Gentile}
\affiliation{CNR-SPIN, Fisciano (Salerno), Italy and Dipartimento
di Fisica "E.R. Caianiello", Universit\`{a} di Salerno, Fisciano
(Salerno)}
\author{Mario~Cuoco}
\affiliation{CNR-SPIN, Fisciano (Salerno), Italy and Dipartimento
di Fisica "E.R. Caianiello", Universit\`{a} di Salerno, Fisciano
(Salerno)}
\author{Dirk~Manske}
\affiliation{Max Planck Institute for Solid State Research,
D-70569 Stuttgart, Germany}

\date{\today}

\begin{abstract}
We study the physical properties of a ballistic heterostructure
made of a ferromagnet (FM) and a spin-triplet superconductor (TSC)
with a layered structure stacking along the direction
perpendicular to the planes where a chiral $p_x+$i$p_y$ pairing
occurs and assuming spin dependent processes at the interface. We
use a self-consistent Bogoliubov-de Gennes approach on a
three-dimensional lattice to obtain the spatial profiles of the
pairing amplitude and the magnetization. We find that, depending
on the strength of the ferromagnetic exchange field, the ground
state of the system can have two distinct configurations with a
parallel or anti-parallel collinearity between the magnetic
moments in the bulk and at the interface. We demonstrate that a
magnetic state having non coplanar interface, bulk and Cooper
pairs spins may be stabilized if the bulk magnetization is assumed
to be fixed along a given direction.
The study of the density of states reveals that the modification
of the electronic spectrum in the FM plays an important role in
the setting of the optimal magnetic configuration. Finally, we find
the existence of induced spin-polarized pair correlations in the FM-TSC system.
\end{abstract}


\maketitle

\section{Introduction}

The past few decades have been marked by a growing interest in the
study of the interplay between superconductivity and
ferromagnetism in heterostructures both for the potential
application in the field of spintronics and because of the
underlying fundamental
physics~\cite{Zutic04,Buzdin05,Bergeret05,Eschrig08}. The control
of the spin degree of freedom in charge currents that can flow
without dissipation represents one of the major challenges in the
context of hybrids based on superconductors and ferromagnets. In
this respect, the presence of spin-triplet pair correlations in
ferromagnet(FM)-superconductor(SC) heterostructures is expected to
play a key role in the generation of new types of spin-valves or
unconventional Josephson junctions.

One of the major obstacle to such achievement is the difficulty in
obtaining a coexistence of superconductivity and ferromagnetism
within a uniform quantum state. The competition between the two
orders can be reduced by a phase or amplitude modulation of the
superconductor order parameter, as for the so called
Fulde-Ferrell-Larkin-Ovchinnikov (FFLO) state~\cite{FFLO}, by
having ferromagnetic domains with no net magnetization in average
on the scale of the superconducting coherence length, or via
specific mechanisms driving the ferromagnetism~\cite{Cuoco2003}.
Such incompatibility, due to the opposite nature of the order
parameters, gets less severe in a ferromagnet- spin-singlet
superconductor (SSC) heterostructure through the occurrence of
nanoscale proximity effects at the interfaces. One remarkable
feature due to the proximity effects is an induced pairing
amplitude in the ferromagnet, characterized by a damped
oscillatory behaviour~\cite{Demler97,Radovic91,Jiang95}. This
leads to a nonmonotonous dependence of the superconducting
critical temperature as well as a 0-$\pi$ transition in the
Josephson current and the possibility to suppress the
superconductivity according to the FM
thickness~\cite{Ryazanov01,Buzdin05,Khusainov97,Halterman04}.
Moreover, spin-triplet pair correlations being even in space and
odd in time or frequency can also be induced in the presence of
magnetic inhomogeneities and significantly contribute in the spin
polarized supercurrent
~\cite{Yokoyama2007,Linder2009a,Linder2010,Bergeret01,Bergeret03,Lofwander05,Halterman07,Asano07,
Braude07}.
Experimental evidence for such odd-in-time spin-triplet pairs was
accomplished either by introducing magnetic inhomogeneity
artificially, or by relying on a source of inhomogeneity intrinsic
to the materials in SC-FM-SC Josephson
junctions~\cite{Keizer2006,Sosnin2006,Khaire2010,Robinson2010,Sprungmann2010,
Anwar2010,Klose2012}. Non uniform magnetic configurations can be
intrinsic or artificially designed, as for instance in the case of
a FM1-FM2-SSC heterostructure with noncollinar magnetization
between FM1 and FM2 ferromagnets. The FM2 ferromagnet acts in this
configuration as a spin-active interface where its magnetization
orientation with respect to the FM1 magnetization is the key
parameter to tune the proximity effects~\cite{Linder2009b}. In turn the presence of
the FM2 layer results into a change of the pair-breaking intensity
and of the superconducting critical temperature. Both the
proximity effects and the induced spin-triplet pairing can
contribute to generate a spin-valve effect by turning the
superconductivity on and off depending on the pattern of the
magnetic profile\cite{f1f2S,f1f2S1,f1f2S2,f1f2S3}.

\begin{figure}
\begin{center}
\includegraphics[scale=0.25]{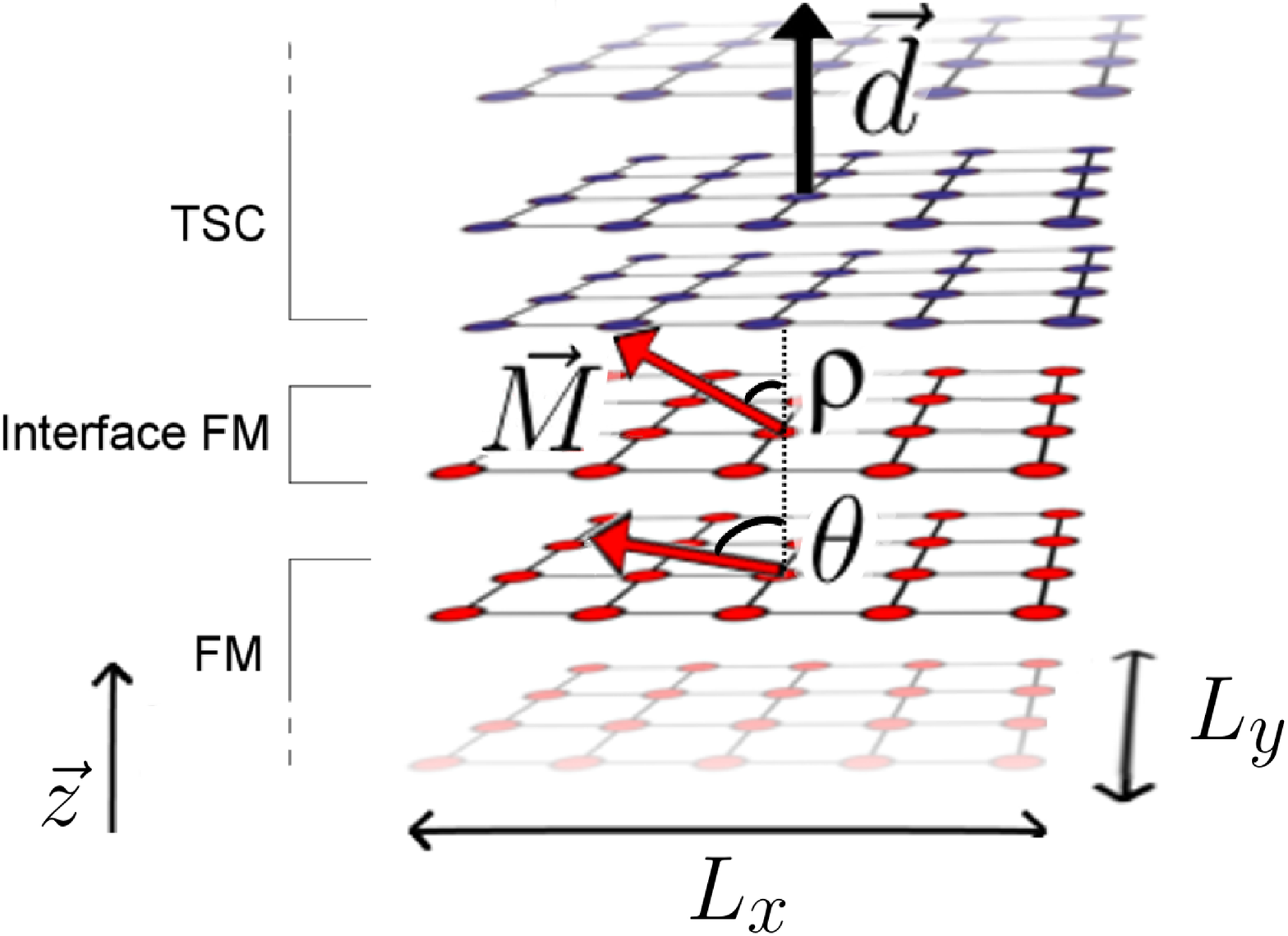}
\caption{(color online) FM-TSC heterostructure with the interface
perpendicular to the $z$-direction. The blue layers are for
the spin-triplet superconductor with a $\vec{d}$-vector (thick
black arrow) perpendicular to the $(xy)$-plane and the red layers
indicate the ferromagnetic region. The red thin
arrow stands for the magnetization $\vec{M}$ in the interior of
the ferromagnetic region (at the interface),
due to the exchange field forming a relative
angle $\theta$ ($\rho$) with respect to the $\vec{d}$-vector. $L_x$ and $L_y$ are the in-plane lateral size of the
heterostructure.} \label{fig.system.001.theta.rho}
\end{center}
\end{figure}

Differently from the induced triplet correlations, a spin-triplet
superconductor (TSC) is marked by a broken symmetry configuration
due to a characteristic pairing potential. Spin-triplet
superconducting order has been predicted in systems like the
heavy-fermion compounds UPt$_3$~\cite{Tou} and
UGe$_2$~\cite{Saxena00}, the quasi-one-dimensional organic
Bechgaard salts (TMTSF)$^2X$~\cite{org}, and the perovskite
ruthenium oxide Sr$_2$RuO$_4$~\cite{Maeno_RMP}. While in the first
two cases a non-unitary pairing seems to occur, a unitary
chiral-state where time-reversal symmetry is broken by orbital
degrees of freedom is realized in the latter~\cite{Luke98}. A
peculiarity of this state is its non-trivial topological
character, specified by a fully open excitation energy gap
coexisting with topologically protected gapless edge states.
Spin-triplet pairing may also manifest itself in
non-centrosymmetric superconductors (NCS)~\cite{Bauer04,Yuan06}
together with the spin-singlet component, due to the lack of a
crystal inversion center that leads to an asymmetric spin-orbit
interaction and in turn results into parity
mixing~\cite{Gorkov01}. The increasing number of materials having
a superconducting order parameter with a spin-triplet symmetry
makes more achievable the design of TSC based heterostructures. In
particular the recent preparation~\cite{Krockenberger2010} of
superconducting thin films of Sr$_2$RuO$_4$~\cite{MacMae2003}
provides a key milestone for the development of such hybrid
systems. Within this context the epitaxial growth of the cubic
itinerant ferromagnet SrRuO$_3$ on the top of Sr$_2$RuO$_4$ could
allow one to obtain an FM-TSC heterostructure with the desired
control of the interface.

The physical behaviour of the FM-TSC heterostructure is expected
to be quite different from that of the FM-SSC. This is especially
due to the basic different structure of the spin-triplet order
parameter that has a vectorial nature and thus can couple to the
orientation of the FM moment as well as to the orbital one. Such
features play a distinct role in controlling the dependence of the
proximity effects in TSC-FM bilayers with diffusive
FM~\cite{Annunziata2011} and the sign of the current in TSC-FM-TSC
Josephson junctions~\cite{TFT}. The interface with the FM also
induces specific effects in the TSC generating a Josephson-like
coupling between the spin $\uparrow$ and $\downarrow$ condensates
of the TSC~\cite{Brydon2009,Brydon2011}. Furthermore, in a FM-TSC
heterostructure with a planar geometry the ferromagnet
magnetization tends to favor a specific orientation in a way that
depends on the orbital symmetry of the order parameter rather than
that of the Cooper pairs spin~\cite{Gentile2012}. The role of
spin-misalignment on hybrids based on TSC, in analogy with the
FM1-FM2-SSC systems, can significantly modify the above scenario
and yield novel physical behaviors. So far it has been considered
only for the case of a half-metal ferromagnet demonstrating that
extra midgap states can be induced as well as a nonzero charge
conductance~\cite{Linder10}.

In this paper we aim to study the role of \textit{spin-dependent}
interface processes in determining the proximity effects and the
physical properties of an FM-TSC heterostructure. In particular,
we consider FM and TSC layers stacked along the $z$-direction
perpendicular to the plane where a chiral $p_x$+i$p_y$ pairing
occurs in the TSC and with a spin-active interface such as to
mimic an FM1-FM2-TSC heterostructure (see
Fig.~\ref{fig.system.001.theta.rho}). We employ a fully
self-consistent Bogoliubov-de Gennes approach on a lattice to
determine the proximity behaviour through the spatial profile of
the spin-triplet pair amplitude and the magnetization.
We obtain the optimal orientation of the
magnetization, for the bulk and the interface, in the presence of
the spin-triplet superconductor. We found that the optimal
configuration is highly depending on the strength of the
ferromagnetic exchange field. While the ground state of the
ferromagnet-normal interface can exhibit two distinct magnetic
configurations, with spin isotropic parallel or antiparallel magnetic
moments among the interface and the interior of the ferromagnet,
the presence of the triplet superconductor favors the
anti-parallel configuration with in-plane magnetization and can
lead to the emergence of a magnetic bulk-interface
noncollinearity. This misalignment tunes the coupling between
ferromagnetism and superconductivity by leading to a
competition between the energy contributions associated with the
proximity effects and with the superconducting order parameter and
those arising from the energy spectrum reconstruction at the interface.
We determine the phase diagram with respect to the most favorable
magnetic orientation and discuss the microscopic mechanisms that
control the settling of the magnetic configuration. Finally, we
discuss how spin-polarized pair correlations are induced in the FM
and at the FM-TSC interface depending on the magnetization
orientation in the ferromagnet with respect to the direction of
the Cooper pairs spin. We believe that the study of such a
geometric configuration has a relevant impact in view of a
possible application to a Sr$_2$RuO$_4$ heterostructure stacked
along the $c$-axis.

The paper is organized as follows. In the Sect.~II we introduce
the model Hamiltonian and the relevant physical quantities to be
determined. The Sect.~III and the Sect.~IV are devoted to
the proximity effects in the FM-TSC and FM$_1$-FM$_2$-TSC systems
respectively. In the Sect.~V we analyze the spin-polarized pair
correlations while a qualitative physical scenario is presented in
the Sect.~VI. Finally, the concluding remarks are
provided in the Sect.~VII.

\section{The model}

We consider a FM-TSC heterostructure with a layered geometry (see
Fig.~\ref{fig.system.001.theta.rho}) described by a single-band
tight-binding model Hamiltonian with an in-plane attractive
nearest-neighbor interaction such as to yield a spin-triplet
pairing with chiral symmetry in the superconducting side. In the
framework of an effective Stoner model for itinerant magnets, an
exchange field $h$ ($h_{{\text{int}}}$) is introduced to yield a
nonzero spin polarization in the ferromagnetic layers (at the
interfacial layer with the TSC). Denoting by $z$ and
$\{x,y\}$ the directions perpendicular and parallel to the
interface layer, respectively, we assume that the system is
uniform along the $x$ and $y$ axes, so that the translational
symmetry is broken only in the $z$ direction. The total
Hamiltonian of the system is then defined on a lattice with size
$L_x\times L_y \times L_z$ ($L_x=L_y=L$) with periodic boundary
conditions along $x$ and $y$. The simulation is performed for a
system size with $L=40$ and $L_z=40$. We indicate each site by a
vector $\vec{i} \equiv(i,i_z)$, with $i$ labeling the $(xy)$
in-plane atomic positions and $i_z$ the different layers along the
$z$-direction. The spin-active layer is placed at the site $i_z=0$
between the FM region ($i_z\leq0$) and the TSC one ($i_z>0$). The
hopping amplitude is given by $t_x=t_y=t$ for the in-plane
processes and $t_z=t/2$ uniformly for all the inter-layers charge
transfer, including that at the FM-TSC interface. A different
choice of the single particle electronic parameters and the system
size alters only quantitatively and not qualitatively the
presented results. All the energy scales are in unit of $t$ and
for clarity, the exchange fields have been scaled with respect to
the amplitude $h_{\text{HM}}=3.15\,t$ corresponding to the
ferromagnetic half-metallic state, i.e. when there is only one
type of carriers at the Fermi level and the minority spin
electrons have a gap in the excitation spectrum. With the above
assumptions the Hamiltonian can be expressed as
\begin{eqnarray}
&H& = \sum_{k,i_{z},\sigma} (\epsilon_{k \sigma} -\mu)
c^{\dagger}_{k \sigma}(i_z) c^{}_{k \sigma}(i_z) +\nonumber
\\&-& \sum_{<i j>, i_z>0 } V \left[ n_{i \uparrow}(i_z)
n_{j \downarrow}(i_z) + n_{i \downarrow}(i_z)
n_{j \uparrow}(i_z) \right] \nonumber \\
&+& \sum_{\delta=\pm1} \sum_{k \sigma} t_z (c^{\dagger}_{k
\sigma}(i_z + \delta) c^{}_{k \sigma}(i_z) + h.c. ) +\nonumber \\
&-& \sum_{i_z<0}
\vec{h}\cdot\vec{s}(i_z)-{\vec{h}_{int}}\cdot\vec{s}(0)
\label{ham}
\end{eqnarray}
with $c_{k\sigma}(i_z)$ being the annihilation operator of an electron
with planar momentum $k$, spin $\sigma$ at the layer $i_z$ and
where $\langle i j\rangle $ denotes the in-plane nearest-neighbor
sites, $\mu$ is the chemical potential and $V$ is the in-plane pairing
strength. Moreover $\vec{s}(i_z) =
\sum_{i,s,s'}c^{\dagger}_{i\,s}(i_z){\vec{\sigma}}_{s,s'}c_{i\,s'}(i_z)$
is the spin density of the layer $i_z$, and $\epsilon_{k
\sigma}=-2 t [\rm{cos}(k_x)+\rm{cos}(k_y)]$ is the in-plane tight-binding
spectrum.\\

For the analysis of the proximity effects and of the most
energetically favorable magnetic configurations the quartic term
of the model Hamiltonian in Eq. \ref{ham} is decoupled in the
Hartree-Fock approximation within the pairing channel. We
introduce the pairing amplitude for an in-plane bond
$F_{ij}(i_z)=\langle c_{i\,\uparrow}(i_z) c_{j\,\downarrow}(i_z)
\rangle$ at a given layer $i_z$. The average $\langle A \rangle$
indicates the thermal expectation value of the operator $A$.
Within such decoupling scheme, the Hartree-Fock Hamiltonian is

\begin{eqnarray}
&H_{HF}& = \sum_{k,i_{z},\sigma} (\epsilon_{k \sigma} -\mu)
c^{\dagger}_{k \sigma}(i_z) c^{}_{k \sigma}(i_z) +\nonumber
\\&+& \sum_{<i j>, i_z>0 } V \left[
F_{ij}(i_z) c^{\dag}_{j \downarrow}(i_z)c^{\dag}_{i \uparrow}(i_z)  \right . \nonumber \\  &+& \left . F_{ji}(i_z) c^{\dag}_{i \downarrow}(i_z)c^{\dag}_{j \uparrow}(i_z) + h.c. \right] \nonumber \\
&+& \sum_{\delta=\pm1} \sum_{k \sigma} t_z (c^{\dagger}_{k
\sigma}(i_z + \delta) c^{}_{k \sigma}(i_z) + h.c. ) +\nonumber \\
&-& \sum_{i_z<0}
\vec{h}\cdot\vec{s}(i_z)-{\vec{h}_{int}}\cdot\vec{s}(0) .
\label{hamHF}
\end{eqnarray}

From the pairing amplitude $F_{ij}(i_z)$ it is possible to build up the
superconducting order parameter with the desired symmetry,
to be then determined self-consistently~\cite{Kuboki2004,Cuoco2008}.
By a proper selection of the
spin-triplet channel one can derive the pairing amplitude for each
layer by summing up the contributions over all the in-plane $k$
vectors.

In general, for a spin-triplet superconductor the order parameter
can be expressed in a matrix form as
\begin{eqnarray*}
\Delta(k)=
\left(\begin{array}{cc}
  \Delta_{\uparrow\uparrow}(k) & \Delta_{\uparrow\downarrow}(k)\\
  \Delta_{\downarrow\uparrow}(k) & \Delta_{\downarrow\downarrow}(k)
\end{array}\right)
= \left(\begin{array}{cc}
  -d_x+i d_y & d_z \\
  d_z & d_x+id_y
\end{array}\right) \, ,
\end{eqnarray*}
\noindent where the $\vec{d}$-vector components are related to the
pair correlations of the various spin-triplet configurations.
Indeed,
$d_x=\frac{1}{2}(-\Delta_{\uparrow\uparrow}(k)+\Delta_{\downarrow\downarrow}(k))$,
$d_y=\frac{1}{2 i
}(\Delta_{\uparrow\uparrow}(k)+\Delta_{\downarrow\downarrow}(k))$,
and $d_z=\Delta_{\uparrow\downarrow}(k)$ are expressed in terms of
the equal spin
$\Delta_{\uparrow\uparrow}(k) ~\mathrm{and}~
\Delta_{\downarrow\downarrow}(k)$, and the anti-aligned spin
$\Delta_{\uparrow\downarrow}(k)$ pair potentials. For the model
upon examination the pairing interaction $V$ is non zero in the
$\uparrow\downarrow$ channel and thus the only non-vanishing order
parameter is $\Delta_{\uparrow\downarrow}(k)$. This implies that
the $\vec{d}$-vector is oriented along the $z$-direction, which is
chosen to be perpendicular to the layers as indicated in
Fig.~\ref{fig.system.001.theta.rho}.

It is worth pointing out that since we are dealing with a hybrid
structure where the superconductor is interfaced to a ferromagnet,
the spin-triplet pair correlations can have also contributions in
the $\uparrow\uparrow$ and $\downarrow\downarrow$ channels. For
this reason it is useful to introduce the following pair
correlators
\begin{eqnarray}
F_{p_{\alpha}}(i_z)&&=\frac{1}{L_x\,L_y} \sum_{k} \rm{sin}(k_{\alpha}) \langle
c_{k\,\uparrow}(i_z) c_{-k\,\downarrow}(i_z) \rangle \, \label{eqspin0} \\
F^{\sigma \sigma}_{p_{\alpha}}(i_z)&&= \frac{1}{L_x\,L_y} \sum_{k}
\rm{sin}(k_{\alpha}) \langle c_{k\,\sigma}(i_z) c_{-k\,\sigma}(i_z) \rangle
\label{eqspin}
\end{eqnarray}
with $\alpha=x,y$  and $\sigma=\uparrow, \downarrow$.\\

The analysis is performed by choosing the chemical
potential ($\mu=-1.8 t$) such that the most stable superconducting
state manifests a chiral $p_x+ip_y$ symmetry. This is a stable state
for the chosen chemical potential and holds for all the values of $V$\cite{Kuboki2001}.
In general, for the examined model Hamiltonian, there is
a competition between the spin-triplet state and
different types of spin-singlet pairing functions, i.e. $d$-wave
and extended $s$-wave, as a function of temperature and chemical
potential. However the chiral spin triplet is more stable than the spin-singlet ones for
a suitable window of chemical potential values, i.e. about
$-1.5<\mu<2.25$. This is because the Fermi surface gets closer to the
points $(\pm \pi/2,\pm \pi/2)$ and therefore the k-dependent pairing
potential, being proportional to (sin($k_x$),sin($k_y$)), can be
maximized there. In order to assess the role of the pairing
strength in the interplay between superconductivity and
ferromagnetism we have also considered different values of the pairing
interaction. We point out that the pairing develops in the $xy$
plane and thus a modification of its strength has a consequence on
the amplitude of the order parameter and the planar coherence length,
while it weakly affects the pair correlations length along the
out-of-plane direction.

Concerning the ferromagnetic side of the heterostructure the inner
layers magnetization $\vec{M}$ depends on the exchange field
${\vec{h}}$, directly proportional to $\vec{M}$, and
its orientation is given by fixing the angle $\theta$ with respect to
the direction of the $\vec{d}$-vector (Fig.~\ref{fig.system.001.theta.rho}).
On the other hand, the
interface exchange field ${\vec{h}_{\text{int}}}$, forming an angle
$\rho$ with respect to the $\vec{d}$-vector, is used to control
the interface magnetic moment
(Fig.~\ref{fig.system.001.theta.rho}). The angle $\theta$ and
$\rho$ are treated as variational parameters. Due to the
rotational spin symmetry of the superconducting state, the $x$-
and $y$- directions are equivalent. Hence, we can consider
${\vec{h}}$ to lie in the ($xz$)-plane with components given by
${\vec{h}}=h\,(\sin(\theta),0, \cos(\theta))$ and similarly for
${\vec{h}_{\text{int}}}=h_{\text{int}}(\sin(\rho),0, \cos(\rho))$.\\

After having diagonalized the Hartree-Fock
Hamiltonian $H_{\text{HF}}$ (Eq.(\ref{hamHF})), the Gibbs free energy can be
determined as
\begin{eqnarray}
G = - \dfrac{1}{\beta} ~ \mathrm{ln} ~ \mathrm{Tr} [
\mathrm{e}^{-\beta H_{\rm{HF}}} ]  \label{eq.G}
\end{eqnarray}
where $\beta=\frac{1}{k_B T}$ is the inverse temperature, $k_B$
being the Boltzmann constant. In order to separate the energy
contributions resulting from the proximity effects and from the
modification of the energy spectrum by those that enters in the
superconducting condensation energy, via the change of the order
parameter in the TSC side of the heterostructure, it is useful to
introduce the quantity $E_{\text{op}}$ expressed as
\begin{eqnarray}
E_{\text{op}} =  - \left(\frac{2}{L_z}\right) \sum_{i_z>0} \left[
|F_{p_x}(i_z)|^2 +|F_{p_y}(i_z)|^2 \right]\,. \label{eq.Eop}
\end{eqnarray}
Such quantity reflects the changes of the superconducting order
parameter in the FM-TSC heterostructure related to the
magnetization configuration of the ferromagnetic layers. In particular,
a minimum in $E_{\text{op}}$ implies a maximum amplitude of the
superconducting order parameter.

Furthermore, for clarity the results are presented by rescaling
the Gibbs free energy and the order-parameter derived energy to
their minimum value. Indeed, we introduce the quantities $\Delta G
= G - G^{\text{min}}$ and $\Delta E_{\text{op}} = E_{\text{op}} -
E_{\text{op}}^{\text{min}}$ where $G^{\text{min}}$ and
$E_{\text{op}}^{\text{min}}$ are the minimum amplitude of the
Gibbs energy and of the order parameter energy, respectively, in the analyzed range of
parameters. This rescaling allows one to directly extract
the energy scale that sets the variation of the Gibbs energy
landscape and to compare different energy profiles as a function
of the physical parameters involved in the analysis. Still, the
variation of $\Delta G$ and $\Delta E_{\text{op}}$ are presented
by scaling their amplitude to the value of $E_{\text{op}}$ associated with a single superconducting layer in the homogenous state.

\section{Proximity effects and optimal magnetic configuration of FM-TSC system}

In this section we present the evolution of the pairing amplitude
of the FM-TSC system assuming that the ferromagnet has a uniform
magnetization i.e., $\theta=\rho$. \\

\begin{figure}
\begin{center}
\includegraphics[scale=0.33]{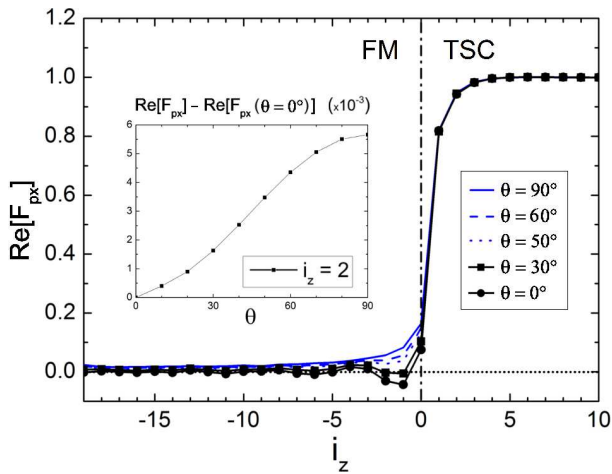}
\caption{(color online) Spatial variation of the real part of the
$p_x$-wave pairing amplitude $F_{p_x}$ scaled to the TSC bulk
value for various angles $\theta$ of the exchange field with
respect to the $\vec{d}$-vector as defined in
Fig.~\ref{fig.system.001.theta.rho}. For this case we assume
$h=h_{\text{int}}=0.16$ and $\theta=\rho$. The FM-TSC interface ($i_z=0$) is
denoted by the dashed line. The inset shows the evolution of the
pairing amplitude $F_{p_x}$ as a function of the angle $\theta$ at
a given layer $i_z=2$.}
\label{fig.op.theta_variations}
\end{center}
\end{figure}

To analyze the most stable magnetic configurations we firstly
study the modification of the pairing amplitude as a function of
the relative angle between the exchange fields and the
$\vec{d}$-vector. As depicted in
Fig.~\ref{fig.op.theta_variations} the spatial dependence of the
real part of the pairing amplitude for the $p_x$ component has a
distinct behavior with respect to the angle $\theta$ when
considering the FM and the TSC side of the heterostructure. The
imaginary part of the $p_y$ component exhibits the same behaviour
of the $p_x$ counterpart due to the tetragonal symmetry of the
system. In the TSC region the pairing amplitude is reduced close
to the interface, with a maximum reduction for $\theta=0$ (see the
inset of Fig. \ref{fig.op.theta_variations}). The main outcome of
this analysis is that the pairing amplitude has a monotonous
dependence on the angle $\theta$ and it reaches its maximum when
the magnetization is aligned in-plane perpendicularly to the
$\vec{d}$-vector.

On the other hand in the FM side the pairing amplitude exhibits a
spatial damped oscillatory behavior at any angle $\theta\neq\pi/2$
and a pure monotonous dependence when the magnetization is
perpendicular to the $\vec{d}$-vector. The oscillatory behaviour
of the pairing amplitude comes from the spin pair-breaking effect
of the ferromagnet. Indeed, the Cooper pairs spin is perpendicular
to the $\vec{d}$-vector, and therefore the only non pair-breaking
configuration for the superconducting components within the
ferromagnet happens when $\theta=\pi/2$.

Moreover, the oscillation length does not depend on $\theta$
because the split of the energy for the up and down electrons,
that controls the modulation in the proximity, is always
proportional to the amplitude of the ferromagnetic exchange field $h$
and not on the pair-breaking projection. It is worth pointing out
that differently from the SSC junction, the oscillation of the
pairing amplitude leads to a negative phase shift only for a value
of $\theta$ in the range $\sim [0,\pi/4]$. Still, we note that in
this case there are no spin singlet induced components at the
interface differently to what is observed in a planar FM-TSC
heterostructures where the interface explicitly breaks the
inversion symmetry in the spatial directions that enter into the
spin-triplet order parameter\cite{Cuoco2008}.\\

\begin{figure}
\begin{center}
\includegraphics[scale=0.34]{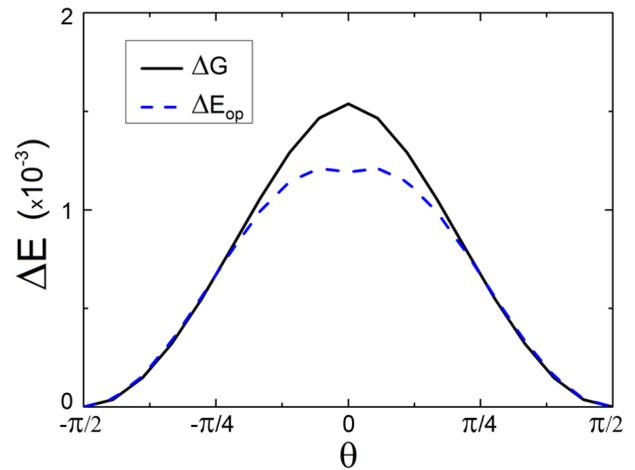}
\caption{(color online) variation of the Gibbs free energy (solid
black line) and the order-parameter derived energy $E_{op}$ (blue
dotted line) as functions of the angle $\theta$ at a given value
of the ferromagnetic exchange $h=0.32$. Both energies are scaled
with respect to their minimum amplitude ($\Delta
G=G-G^{\text{min}}$ and $\Delta
E_{op}=E_{op}-E_{op}^{\text{min}}$) and are renormalized to the
$E_{\text{op}}$ value of a single superconducting layer in the
homogenous state.}
\label{fig.energies.theta_variations}
\end{center}
\end{figure}

We are now interested in the energetically most favorable
magnetization orientation in the ferromagnetic layers. We have
analyzed the evolution of the Gibbs free energy $G$ and of the
energy associated with the TSC order parameter $E_{{\text{op}}}$
for different values of the relative orientation of the
ferromagnetic moment with respect to the $\vec{d}$-vector and as a
function of the strength of the ferromagnetic exchange field.

As one can note in the Fig.~\ref{fig.energies.theta_variations} the
Gibbs free energy has a non monotonous profile with two degenerate
minima at $\theta=\pm \pi/2$ and a maximum for $\theta=0$. This
means that, due to the coupling with the TSC, the energetically
most favorable magnetic configuration for the ferromagnet is
achieved when the magnetization lies in the plane of the Cooper
pairs spin. Otherwise, the maximal loss in the energy occurs when
the magnetization is parallel to the $\vec{d}$-vector. To
understand the driving force in the stability of the magnetic
configuration it is instructive to analyze the behavior of the
order-parameter derived energy $E_{\text{op}}$. The results
indicate that the maximum energy gain due to the variation of the
superconducting order parameter in the TSC region occurs for a
magnetization that lies perpendicular to the $\vec{d}$-vector
which also corresponds to a lack of spin-pair breaking in the
proximity effect. Since the total energy has the same minima than
the $E_{\text{op}}$ term one can deduce that for this strength of
the exchange field the changes of the spectra in the Gibbs free
energy due to the proximity effect and to a possible electronic
reconstruction at the interface cooperate with the gain in the
condensation energy to stabilize the observed magnetic
configuration.

The trend observed for the pairing amplitude as a
function of the angle $\theta$ can be discussed in the framework
of the scattering formulation for the TSC-FM heterostructure. The
pairing amplitude variation is mainly controlled by the Andreev
reflections at the interface. Since Andreev scattered electron
(hole) leads to hole (electron) pairs formation in the
superconductor such process is expected to drive the change of the
pairing order parameter close to the interface. The resulting
interference of the different scattering amplitude is maximal when
the Andreev reflected Cooper pair acquires a spin and/or an
orbital phase
shift~\cite{gapsuppress,Brydon2009,Brydon2011,Gentile2012}. This
observation can be justified by the fact that
 a sign change can lead to
destructive interference. In the case upon examination in the
paper, the orbital phase change for the given
heterostructure is 0 at any angle of incidence of the
electron that undergoes an Andreev reflection
and therefore the overall phase shift is
minimized when
the magnetization forms an angle $\pi/2$ with the
$\vec{d}$-vector\cite{Brydon2009}.

\section{Spin-active processes at FM-TSC interface}

In this section we analyze the role of
spin-dependent processes at the FM-TSC interface with the aim to
find the most favorable magnetic orientation in the presence of an
extra degree of freedom for the spin moments at the interface.
Concerning the modification of the pairing amplitude in the TSC,
the presence of a spin-scattering interface does not modify
substantially the trend observed for the uniform case with
$\theta=\rho$.  Indeed, the TSC order parameter is maximized when
both the magnetization at the interface and in the bulk lies in
the plane perpendicular to the $\vec{d}$-vector. More
specifically, when restricting the magnetic orientation to the
$xy$-plane, we find that the pairing amplitude in the TSC region
increases for $\theta$ and $\rho$ being collinear and
antiparallel. The behaviour in the FM region shows that the
intensity of the paring amplitude has an oscillatory/monotonous
profile similar to the case with uniform magnetization.
Nevertheless, for the planar orientations it turns out to be
larger for the parallel interface-bulk configuration if compared
to the antiparallel one.
This implies a reduction of the proximity effect when the bulk-interface
magnetic moments are antiparallel aligned.\\

\begin{figure}
\begin{center}
\includegraphics[scale=0.29]{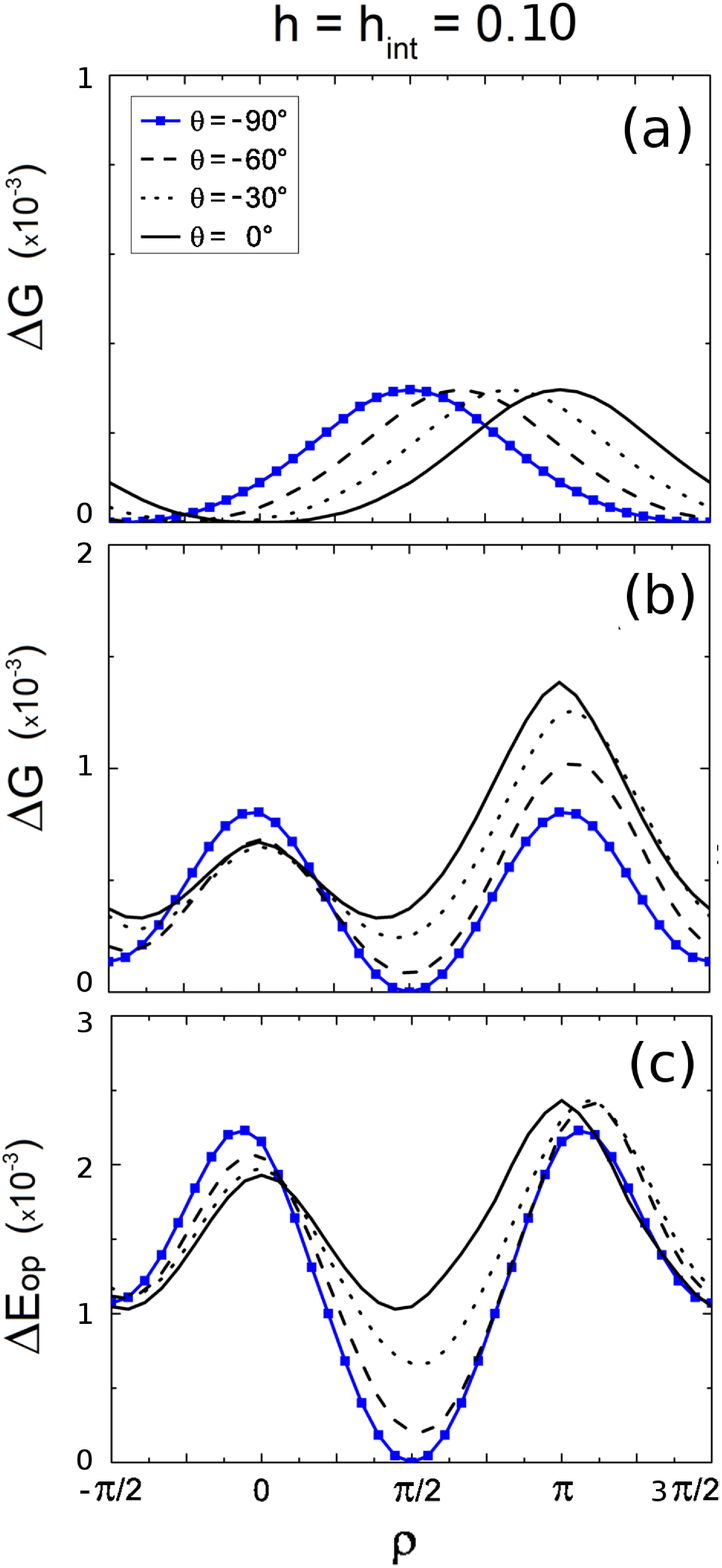}
\caption{(color online) Variation of the Gibbs free energy for V=0.0t (a) and V=-2.0t (b), scaled with respect to its minimum amplitude $G^{\text{min}}$
($\Delta G=G-G^{\text{min}}$) and renormalized to the energy
$E_{\text{op}}$ associated with a single superconducting layer in
the homogenous case, as a function of the angle $\rho$ at various
angle $\theta$ and for a given value of the ferromagnetic exchange field
$h=h_{\text{int}}=0.1$. c) Variation of $E_{\text{op}}$ ($\Delta
E_{\text{op}}=E_{\text{op}}-E_{\text{op}}^{\text{min}}$) energy
associated with the superconducting order parameter in the TSC
region with the same renormalization and parameters as in a).}
\label{fig.energie.all.theta.rho.03}
\end{center}
\end{figure}
\begin{figure}
\begin{center}
\includegraphics[scale=0.29]{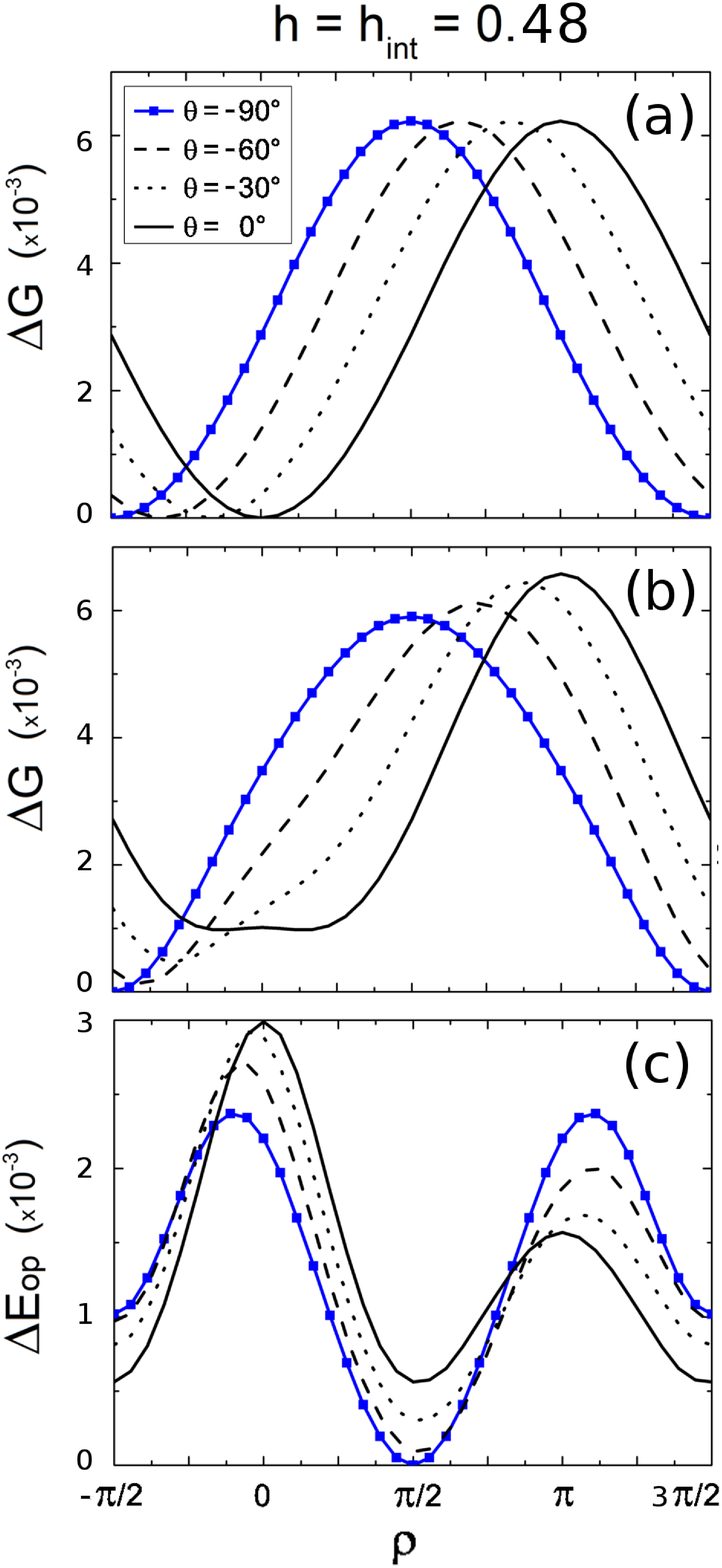}
\caption{(color online) Variation of the Gibbs free energy for V=0.0t (a) and V=-2.0t (b), scaled with respect to its minimum amplitude $G^{\text{min}}$
($\Delta G=G-G^{\text{min}}$) and renormalized to the energy
$E_{\text{op}}$ associated with a single superconducting layer in
the homogenous case, as a function of the angle $\rho$ at various
angle $\theta$ and for a given value of the ferromagnetic exchange field
$h=h_{\text{int}}=0.48$. c) Variation of $E_{\text{op}}$ ($\Delta
E_{\text{op}}=E_{\text{op}}-E_{\text{op}}^{\text{min}}$) energy
associated with the superconducting order parameter in the TSC
region with the same renormalization and parameters as in a).}
\label{fig.energie.all.theta.rho.10}
\end{center}
\end{figure}

\subsection{Optimal magnetic configuration}

\begin{figure}
\begin{center}
\includegraphics[scale=0.29]{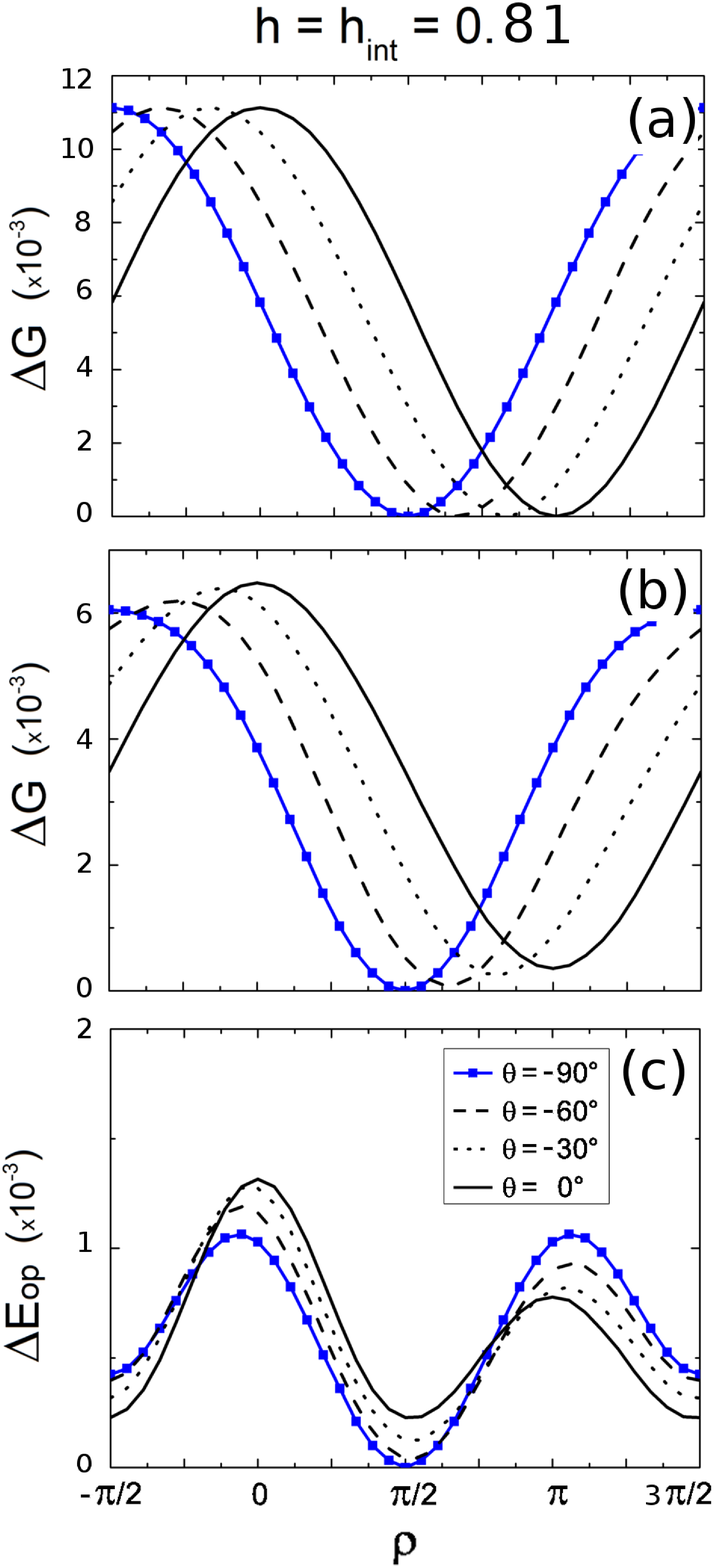}
\caption{(color online) Variation of the Gibbs free energy for
V=0.0 (a) and V=-2.0t (b), scaled with respect to its minimum
amplitude $G^{\text{min}}$ ($\Delta G=G-G^{\text{min}}$) and
renormalized to the energy $E_{\text{op}}$ associated with a
single superconducting layer in the homogenous case, as a function
of the angle $\rho$ at various angles $\theta$ and for a given
value of the ferromagnetic exchange field $h=h_{\text{int}}=0.81$. c)
Variation of $E_{\text{op}}$ ($\Delta
E_{\text{op}}=E_{\text{op}}-E_{\text{op}}^{\text{min}}$) energy
associated with the superconducting order parameter in the TSC
region with the same renormalization and parameters as in a).}
\label{fig.energie.all.theta.rho.20}
\end{center}
\end{figure}

The analysis of the energies variations reveals a
complex scenario because the presence of spin-dependent processes
at the interface tends to modify the balance in the
Gibbs free energy due to the electronic spectrum and to the
superconducting order parameter. In the following we consider two
possible physical situations for the search of the most favorable
magnetic configuration: i) the magnetization in the bulk and at
the interface of the FM have no preferential orientation (i.e.
$\theta$ and $\rho$ are treated as independent variational
parameters), ii) the magnetization in the bulk is fixed in
amplitude and orientation (for instance due to an intrinsic source
of magnetic anisotropy or is pinned externally along a specific
direction via an exchange-bias coupling) and we look for the
optimal orientation of the magnetization at the interface which
minimizes the Gibbs free energy. For both cases the amplitude of
the spin moments is chosen by fixing the value of the
ferromagnetic exchange field $h$ and $h_{\text{int}}$ in the bulk
FM layers and at the interface, respectively. The behavior of the
Gibbs free energy $G$ and the order-parameter derived energy
$E_{\text{op}}$ with respect to the angles $\rho$ and $\theta$ are
reported in the
Figs.~\ref{fig.energie.all.theta.rho.03},~\ref{fig.energie.all.theta.rho.10}
and~\ref{fig.energie.all.theta.rho.20} for which $h=h_{\rm{int}}$=0.10,
0.48, and 0.81, respectively, are representative values for the
regimes of weak, intermediate and strong ferromagnet. The
amplitude of the exchange fields are scaled in such a way that
$h=1$ corresponds to the the transition into the half-metallic
state with an energy gap in the minority spin band. Furthermore,
in order to single out the role of the superconductivity in the
energy competition, we have also analyzed the ferromagnet-normal
metal (FM-N) heterostructure assuming the same microscopic
descriptions of the FM-TSC system with zero pairing interaction
$V$.

\begin{figure}
\begin{center}
\includegraphics[scale=0.29]{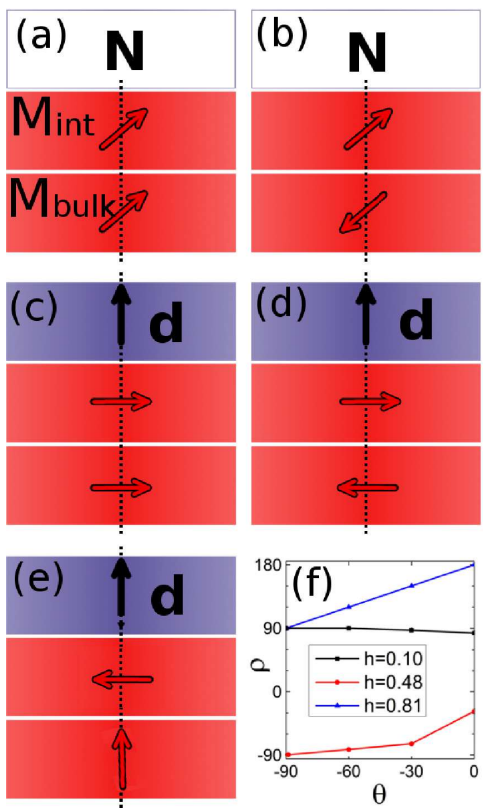}
\caption{(color online) Isotropic parallel (a) and antiparallel (b) configurations
for a FM/N interface. Parallel (c) and antiparallel (d) configurations
in the \textit{xy} plane for a
FM/TSC interface. e) Configuration when the magnetization in the bulk is
fixed out of plane. f) Magnetization angle at the interface $\rho_{min}$
which minimize the free energy for a fixed magnetization angle $\theta$ in
the bulk.}
\label{fig.magnetization.configurations}
\end{center}
\end{figure}

Let us start by the FM-N heterostructure. The
evolution of the ferromagnetic state can be monitored in the Figs.
4(a), 5(a) and 6(a). For the weak and intermediate ferromagnetic
cases, i.e. Figs. 4(a) and 5(a), the minimum of the Gibbs
potential corresponds with the configuration where the
magnetization in the bulk is parallel to the one at the interface
(i.e. $\rho=\theta$, Fig.~\ref{fig.magnetization.configurations}~(a)).
Since we are describing a ferromagnet
without spin anisotropy one finds, as expected, a complete
degeneracy in energy for all the possible spin orientations, i.e.
the ground state, as expected, is rotational invariant in the spin
space. The increase of the ferromagnetic exchange field towards the
half-metallic state brings the ground state to have a transition
into a different configuration. As one can note by inspection of
the Fig. 6(a), the minimum of the Gibbs energy occurs at angles
where the magnetization in the bulk and at the interface are
antiparallel aligned (i.e. $\rho=-\theta$, Fig.~\ref{fig.magnetization.configurations}~(b)).
As for the previous
cases the spin invariance is kept and we find that the ground
state is degenerate at all the possible orientations of the
ferromagnetic magnetization.

When considering the physical case i) where all the magnetic
orientations are assumed to be equivalent within the FM, the
presence of the spin-triplet superconductor can lead to a breaking
of the rotational symmetry for both the magnetic state in the bulk
and at the interface. In this circumstance, since $\theta$ and
$\rho$ are taken as independent variational parameters, it is the
absolute minimum of the Gibbs free energy to indicate the most
favorable magnetic configuration for the FM-TSC system in the
presence of spin-dependent processes at the interface. By
inspection of Figs. 4(b) and 5(b) one can observe that for
$h=h_{\text{int}}=0.10$ and $0.48$ the ground state corresponds to a
configuration with parallel interface-bulk magnetic moments, i.e.
$\theta=\pi/2$ and $\rho=\pi/2$, with the magnetization lying in
the $(xy)$-plane perpendicular to the $\vec{d}$-vector
(Fig.~\ref{fig.magnetization.configurations}~(c)). However,
for $h=h_{\text{int}}=0.81$
(Fig.~\ref{fig.energie.all.theta.rho.20}~(b)) the Gibbs free
energy is minimized by an antiparallel configuration within the
$(xy)$-plane i.e., $\theta=\pi/2$ and $\rho=-\pi/2$
(Fig.~\ref{fig.magnetization.configurations}~(d)). The behavior
of the order-parameter derived energy, as presented in Figs. 4(c), 5(c)
and 6(c), shows that the maximum gain
in $E_{\text{op}}$ always
occurs for an antiparallel interface-bulk configuration.

Moving to the physical case ii) where the exchange field orientation
in the bulk FM is kept fixed while it can vary at the interface,
all the curves of the Gibbs free energy as a function of $\rho$
at a given angle $\theta$ have to be considered separately in order
to determine the ground state configuration.
The resulting outcome is that for weak and intermediate ferromagnets
the minimum of the Gibbs energy is not anymore achieved
when the magnetization in the bulk and at the interface are aligned,
while an antiparallel configuration remains favorable
for any angle $\theta$ in the case of a strong ferromagnet.
The magnetic configuration at the interface is therefore
strongly related to the strength of the ferromagnetic exchange field (see
Fig.~\ref{fig.magnetization.configurations}~(f)).
\begin{figure*}
\begin{center}
 \includegraphics[scale=0.32]{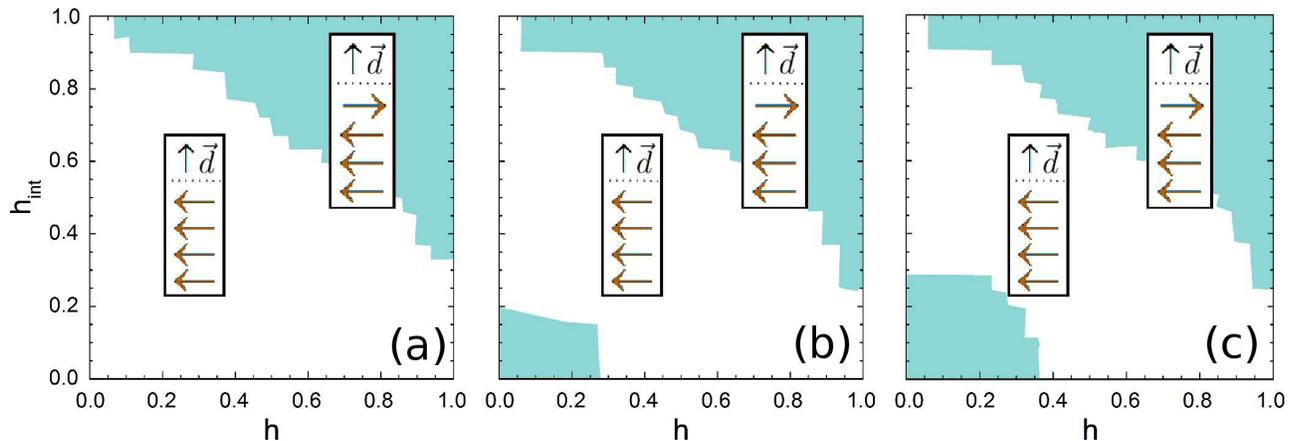}
\caption{(color online) phase diagrams for the energetically most
favorable configuration in the plane $(h_{\text{int}},h)$ for V$=0.0t$ (a), V$=-2.0t$ (b) and
V$=-2.5t$ (c). The
ferromagnetic exchange fields are expressed in units of the
exchange amplitude $h_{\rm{HM}}\cong3.15$ corresponding to the
transition value to an half-metal FM, where namely the density of the
spin minority carriers goes to zero. The sketch represents the
most favorable configurations with parallel (white area) and
anti-parallel (blue area) interface-bulk magnetic moments
alignment, respectively.} \label{fig.phase.diagram}
\end{center}
\end{figure*}
More precisely, the most favorable configuration for a weak
ferromagnet is with a magnetic orientation at the interface that
lies in the plane perpendicular to the $\vec{d}$-vector almost
independently of the magnetic configuration in the bulk. The
Fig.~\ref{fig.magnetization.configurations}~(e) represents for
example the case where the magnetization in the bulk is pinned
perpendicular to the plane, i.e. $\theta=0$. This means that the
superconducting correlations dominate in the scattering processes
and in the energy balance. By increasing the amplitude of the
ferromagnetic exchange field, the correlation between the magnetization
at the interface and that in the bulk is modified in a way that
the resulting ground state has a non collinear magnetic profile.
The optimal magnetic orientation at the interface clearly exhibits
a competition between two effects. We anticipate that this is due
to the competition between scattering processes
associated with the {\it normal} and the Andreev reflected
electrons. On one side the scattering of the electrons with
energies above the superconducting gap would tend to align the
bulk and interface magnetization in a parallel configuration, on
the other hand the Andreev reflected electrons occurring at midgap
energies favor an antiparallel magnetic state with planar spin
orientation. The resulting effect is to develop a magnetic
interface state with components both along the bulk spin direction
and in the plane perpendicular to the $\vec{d}$-vector. For the
considered cases
 $h=0.48$ and $0.82$, the deviation from having $\rho=\theta$ is
due to the presence of the superconducting processes which always
prefers to stabilize an in-plane magnetic orientation.

Indeed, if we look more specifically at the case $h=0.48$,
the optimal orientation at the
interface is at an angle $\rho\approx 3\pi/8$ when the
bulk magnetization is parallel to the $\vec{d}$-vector (i.e. $\theta=0$).
The angle $\rho$
corresponding to the relative minimum of the Gibbs free energy
moves from about $3\pi/8$ to $\pi/2$ when $\theta$ varies from $0$
to $\pi/2$ (see Fig.~\ref{fig.magnetization.configurations}~(f)).
Interestingly, the minimum of the order parameter
energy $E_{\text{op}}$, corresponding to the maximal variation of
the gap amplitude, occurs always at $\rho=-\pi/2$ for any value of
$\theta$ if we exclude the $\theta=0$ configuration where $\pm
\pi/2$ are degenerate in energy due to symmetry reasons. The
comparison between the variation of the Gibbs energy $\Delta G$
and the order parameter energy $\Delta E_{op}$
clearly indicates a competition between a contribution due to the
change of the order parameter in the TSC region, and consequently
of the condensation energy, and another one related to the
magnetic energy close to the interface.

\subsection{Phase Diagram}

The analysis of the Gibbs energy profile indicates that the ground
state is not unique but depends on the strength of the exchange
field. Therefore it is interesting to calculate the complete phase
diagram in terms of $h$ and $h_{\text{int}}$ by evaluating the
absolute minimum of the Gibbs free energy in the $[\theta,\rho]$
landscape.  Furthermore, in order to assess the role of the
superconducting correlations strength in determining the most
stable magnetic configuration, we have also computed the phase
diagram at different values of the attractive interaction for the
TSC.

The results are reported in
Fig.~\ref{fig.phase.diagram}. They show that the parallel and
antiparallel interface-bulk alignment of the magnetization lying
on the $(xy)$-plane are the two most stable configurations. The
antiparallel state is favored for weak and strong ferromagnet
while the parallel one is preferred at intermediate exchange
values. A first order transition occurs at the boundary between
the two regions of stability. Another consequence of the performed
analysis is that the extension of the region for the antiparallel
stability depends on the strength of the superconducting pairing
coupling. This finding underlines the role of the superconducting
state in modifying the magnetic configuration of the ferromagnet.
Finally the boundaries of the antiparallel bulk-interface
configuration for strong ferromagnet regime
are only slightly affected in the direction of shrinking the region of
stability for the parallel magnetic state. Indeed, in this case
the antiparallel configuration is already preferred
without the superconductivity and the ferromagnetic exchange field
is dominant in the proximity effect.

It is worth pointing out that there is an asymmetry in the role
played by the interface and the bulk exchange field which can be
emphasized by scanning the phase diagram along lines with given
$h_{\text{int}}$ and $h$. Indeed at small fixed values of
$h_{\text{int}}$ ($\lesssim$ 0.2) there is no reentrant behavior
as a function of $h$ while one does find the double transition for
the antiparallel configuration in the case of small fixed values
of $h$ by tuning the amplitude of $h_{\text{int}}$. Such asymmetry
underlines the role of the interfacial spin-dependent scattering
in driving the reentrant behavior from the parallel to the
antiparallel configuration in the regime where the magnetization
at the interface is large with a small density of spin-minority
electrons.

\subsection{Density of States}

To get a deeper understanding of the modification on the energy
spectrum induced by the interface-bulk misalignment we evaluate
the density of states (DoS) $\rho(i_z,\omega)$ at each layer
within the FM and TSC region of the heterostructure. To do that,
we analyze the integrated quantity $I(i_z)=\int_{-D}^{0} \omega
\rho(i_z,\omega) d\omega$, representing the contribution to the
Gibbs energy coming from the electronic spectrum, with $- D$ being
the lowest occupied energy level and $0$
the effective Fermi energy for the system after the Bogoliubov rotation.\\

\begin{figure}
\begin{center}
 \includegraphics[scale=0.33]{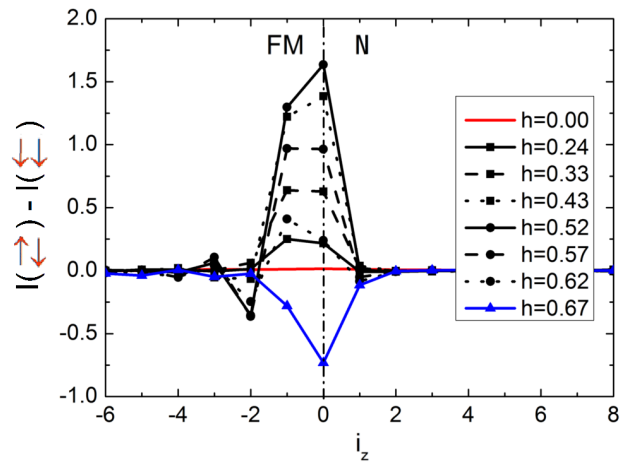}
\caption{(color online) difference of the integrated density of
states for the antiparallel and the parallel configuration as a
function of the layer position and of the exchange field strength for the FM-N interface. We
consider an equal amplitude of the exchange field at the
interface and in the bulk, i.e. $h=h_{\text{int}}$.}
\label{fig.integrated.dos}
\end{center}
\end{figure}

In Fig.~\ref{fig.integrated.dos} we report the spatial evolution
of the difference $\Delta I=I(\begin{array}{c}
\rightarrow \\
\leftarrow
\end{array})-I(\begin{array}{c}
\rightarrow \\
\rightarrow
\end{array})$ between the values of $I(i_z)$, corresponding to the
antiparallel (i.e. $\rho=\pm\pi/2$ and $\theta=\mp\pi/2$) and the
parallel (i.e. $\rho=\theta=\pi/2$) configurations, as a function
of the ferromagnetic exchange field, assuming that $h=h_{\text{int}}$.
Since $I(i_z)$ is a negative quantity, a positive $\Delta I$
indicates that the electronic spectrum contribution tends to favor
the parallel configuration and the opposite for negative
amplitudes. The analysis of the DoS reveals a significant
dependence of the contribution $I(i_z)$ on the magnetic
configuration and on the strength of the ferromagnetic exchange field.
As shown in Fig.~\ref{fig.integrated.dos}, the main change in the DoS occurs at the FM
layers which are close to the interface. This reflects the scale
length where the energy spectrum reconstruction mainly occurs.\\

In the TSC region the electronic reconstruction always favors the
antiparallel state, as it is also expected from the evaluation of
the order parameter energy $E_{\text{op}}$. Thus, no significant
renormalization of the DoS close to the Fermi energy happens that
interferes with the role of the order parameter in the energy
balance.

In the FM side the energy contribution due to the electronic
spectrum is highly nontrivial. For small values of the
ferromagnetic exchange field there is an appreciable difference between
the energy related to the electronic spectrum of the two magnetic
configurations. Such difference becomes larger and larger in favor
of the parallel configuration when $h$ increases from 0.32 to
0.51. Then the difference decreases until the antiparallel
configuration starts to dominate. The results of the DoS confirm a
scenario of electronic modification close to the interface playing
a significant role in the magnetic configuration transition.

It is worth pointing out that the modifications of the DoS are
substantial both in the energy range of the superconducting gap
and also outside it. This is partially due to the superconducting
correlations and in part to the change of the DoS due to the
processes of scattering at the interface at energies which are
larger than the superconducting gap in the presence of exchange field
split Fermi surfaces. Interestingly, in the regime of intermediate
ferromagnetic exchange field, i.e. for $h$ in the range of 0.32 to
0.57, the density of states of the ferromagnetic layers close to
the interface exhibits new electronic states close to the Fermi
energy when the antiparallel configuration is considered. Although
such states may represent a distinctive feature of the FM-TSC
heterostructure, they do not contribute significantly in
determining the stability of the ferromagnetic profile.

\section{Spin-polarized pair correlations}

\begin{figure}
\begin{center}
 \includegraphics[scale=0.35]{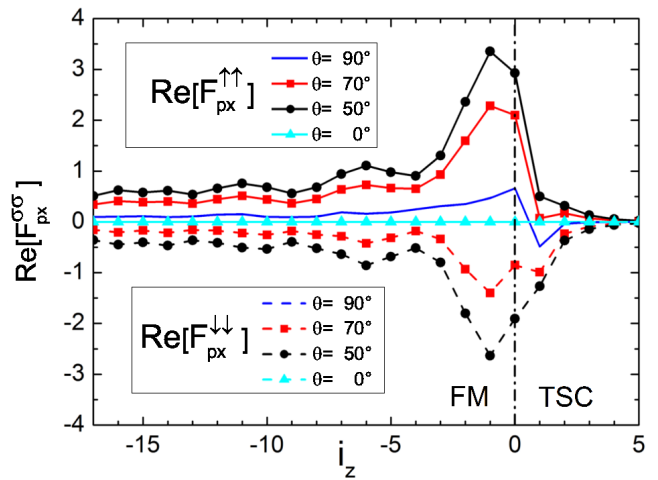}
\caption{(color online) Spatial variation of the real part of
$F^{\uparrow \uparrow}_{p_x}$ and $F^{\downarrow
\downarrow}_{p_x}$ pair correlations associated with the
$\uparrow\uparrow$ (solid lines) and $\downarrow\downarrow$ (dashed line) spin-triplet
configurations for a
representative case of $p_x$ orbital symmetry as a function of the
magnetization direction $\theta$ and at a given exchange field
$h=0.16$. The amplitude of the correlator is scaled to the value of
the bulk one Re$[ F_{p_x} ]$ in the $\uparrow\,\downarrow$ spin
configuration. The FM-TSC interface is denoted by the dashed
line.} \label{fig.op.up.down}
\end{center}
\end{figure}

In this section we analyze the induced spin-triplet pair
correlations, as defined in Eq. \ref{eqspin}, with $p_x$ and $p_y$
orbital symmetry but having parallel $\uparrow\,\uparrow$ and
$\downarrow\downarrow$ spin configuration along the $z$ direction
as compared to the state with zero spin projection associated with
the bulk order parameter, i.e. the superposition
$\uparrow\,\downarrow$+$\downarrow\uparrow$. These pair
correlations are identically zero in the bulk, since the
superconductor has no net spin polarization. However, it is
possible to induce spin polarized pair correlations close to the
interface, since a magnetization that is noncollinear to the
$\vec{d}$-vector can lead to spin-flip processes and, then, to a
consequent spin-mixing. It is instructive to consider the case of
an uniform ferromagnet i.e., $\rho=\theta$, with a magnetization
forming an angle $\theta$ with respect to the $\vec{d}$-vector. We
show in Fig.~\ref{fig.op.up.down} the spatial profile of the
$F^{\sigma\sigma}_{p_x }(i_z)$ $z-$spin polarized pair
correlations for both cases $\sigma=(\uparrow,\downarrow)$ and for
different orientations of the magnetization.\\

\begin{figure}
\begin{center}
 \includegraphics[scale=0.45]{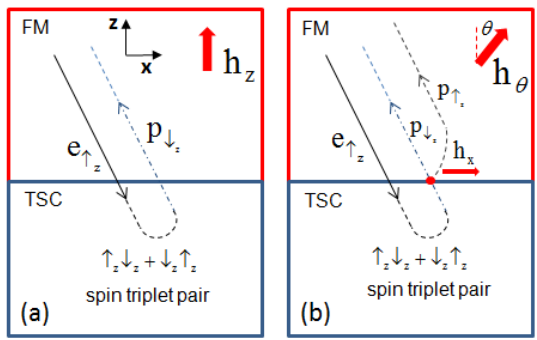}
\caption{(color online) Sketch of the Andreev processes that may
contribute to the build up of the $F^{\uparrow \uparrow}$ and
$F^{\downarrow \downarrow}$ $z$-spin polarized pair correlations.
An electron ($e_{\uparrow_z}$) with spin up aligned along the $z$
direction can be scattered as a hole ($p_{\downarrow_z}$) due to
the spin symmetry of the triplet order parameter. (a) For a
one-component exchange field which is parallel to the
$z$-direction, the only allowed Andreev scattering is to convert
an electron $e_{\uparrow_z}$ into a hole $p_{\downarrow_z}$ with
opposite spin orientation. (b) For an exchange field aligned along
a given direction $\theta$ (the magnetization has component along
$x$ and $z$) it is possible to inject an electron $e_{\uparrow_z}$
with spin up along $z$ and to get an Andreev reflected hole with
opposite spin $p_{\downarrow_z}$ or with the same spin
polarization of the incident particle $p_{\uparrow_z}$ due to the
effect of the transverse $h_x$ component to yield a spin-flip.
This process allows to have $z$-spin polarized pair correlations
along the $z$ direction.} \label{fig.op.up.down.scattering}
\end{center}
\end{figure}

As one can note, a magnetization parallel to the
$\vec{d}$-vector ($\theta=0$) does not induce $z$-spin polarized
pair correlations. This can be understood by observing that to
contribute in the $F^{\sigma\sigma}_{p_x }(i_z)$ amplitude, the
incident electrons and the scattered holes have to be spin polarized along the same
$z$-direction and this can occur only when the Andreev reflection
involves a change in the spin orientation because of the spin
symmetry of the superconducting order parameter, as depicted in
Fig.~\ref{fig.op.up.down.scattering}(a). Therefore, the resulting pair correlations have no
contributions in the $z$-spin polarized pairing channel.

By changing the exchange field orientation from the direction
perpendicular to the ($xy$)-plane (i.e. $\theta=0$) to the
in-plane ($\theta=\pi/2$), the amplitude of the $z$-spin polarized
pair correlations exhibits a nonmonotonous behavior. They increase
upon a critical angle $\theta_{max}\sim \pi/4$ and then decrease
when the exchange field tends to point within the $(xy)$-plane.
The appearance of nonzero correlations for the
$z$-spin polarized pairs when the exchange field is not parallel
to the $\vec{d}$-vector can be addressed by analyzing the process
depicted in Fig.~\ref{fig.op.up.down.scattering}(b). The noncollinearity between the exchange
field and the spin polarization of the incident electrons (holes)
allows for a spin flip of the Andreev reflected holes (electrons) and
therefore generates electron (hole) pairs with parallel spin
orientation along the $z$- direction. More specifically, we note
that the amplitude of the $z$-spin polarized pair correlations is
maximum for angles close to $\theta\sim\pi/4$. Indeed, those
orientations optimize the number of incident
particles with spin polarization parallel to $z$ together with the
exchange amplitude for the mechanism of spin-flip process (i.e.
the $h_x$ component of the exchange field). On the other hand, the
pairing amplitude is minimum when the magnetization lies in the
plane, i.e. at $\theta=\pi/2$. This is because, in this case,
although the amplitude of the exchange field is maximally active
as a spin flip generator, the number of incident particles with
spin polarization parallel to
$z$ is minimum.\\

We also find that the evolution of the pair correlations in the FM
side is decaying and oscillating as a function of the layer
position with the same period of those associated with the bulk
order parameter (see for comparison the Fig.
\ref{fig.op.theta_variations}). Since there is always an energy
split between the Fermi energies of electrons with spin up and
down polarization at any orientation of the magnetization, we do
expect a spatial dependence with an oscillating behavior.
Indeed, the $z$-spin polarized pairs along the $z$
direction get experience of the transverse component of the
exchange field that is nonvanishing at any angle $\theta$ which is
different from zero. Such component acts as a pair breaking term and
makes the pairs acquiring a
finite momentum which results in a spatial modulation of the
pairing amplitude. Moreover, it is interesting to see that in
the TSC side $z$-spin polarized pairs are also induced by a sort
of inverse proximity effect, i.e. due to the penetration of the
magnetization in the superconducting layers. Therefore, in the
case of the existence of a pairing potential component $V^{\sigma
\sigma}$, there would be a rotation of the $\vec{d}$-vector that
can have significant consequences on the electronic transport in a
Josephson junction configuration \cite{Brydon2011}. Still, the
modulation of the pair correlator does not lead to a phase change
as a function of the spatial coordinate.

A change in the amplitude of the ferromagnetic
exchange field does not modify the general trend obtained for the
case of $h=0.16$. In particular, by increasing the ferromagnetic
exchange one gets a reduction of the $z$-spin polarized triplet
correlations. This is because the overall amplitude is set by the
magnitude of the correlator associated with the superconducting
order parameter. Taking into account the mechanisms presented in
Fig.~\ref{fig.op.up.down.scattering}(b) one can also argue that the presence
of a spin-flip process at the interface allows to generate
z-spin polarized pairs even for the case of a bulk
magnetization parallel to the $\vec{d}$-vector. For the other cases, we might
expect that the interface-bulk misalignment tends to modify the
amplitude of the $z$-spin polarized pair correlations close to the
interface in a way that depends on the balance between the number
of incident electrons and the effectiveness of the spin-flip
generator. It is worth pointing out that, since the maximum
amplitude of the $z$-spin polarized triplet correlations
corresponds to an angle $\theta_{max}\sim \pi/4$, they do not have
a relevant role in the balance of Gibbs free energy related to the
proximity effect. On the other hand, in the case of an induced
pair potential in the TSC region in the spin-polarize channel, due
to a modification of the pairing glue at the interface, there
would be a significant change of the energy balance related to the
condensation energy.

\section{Discussion}

In this section we would like to present a qualitative physical
scenario related to the origin of the competing ground states in
the FM-TSC heterostructure in the presence of spin-active
interface processes.\\

We firstly consider the origin of the competition between the
parallel and the antiparallel magnetic configurations within the
FM-N heterostructure. To such purpose, we provide a simplified
picture to justify the stability of a parallel bulk-interface
state in the weak-intermediate ferromagnet and of an antiparallel
one in the strong ferromagnet regime.

In the latter case,  it is useful to consider the modification of the wave function in the
ferromagnetic region close to the interface. We can schematically
discuss this physical situation by means of a scattering
formulation. Assuming that the interface plays the role of a
spin-dependent barrier it is possible to determine the probability
of an electron, propagating through the bulk with a given
spin-direction and momentum, to be reflected or transmitted across
the barrier. By solving the scattering problem, as expected the
probability of reflection of a scattering electron increases as a
function of the barrier potential strength $Z$.

Hence, an incident electron in the spin-majority band of the FM
has experience of a larger (smaller) potential barrier if the
magnetization at the interface is not parallel to its spin, which leads
to a smaller (larger) spin-filtering effect. Still, for the antiparallel
configuration, the spin-minority electrons, having a
spin-polarization parallel to the magnetization at interface, have
a higher probability to be transmitted. The resulting overall
effect is to have more electrons reflected in the spin-majority
band as well as more electrons transmitted in the spin-minority
band, respectively. Thus, there is an increase of the spin
polarization close to the interface because the wave-function has
a larger (smaller) density probability for the spin majority
electrons in the antiparallel (parallel) bulk-interface magnetic
state. Such an increase of the spin-density would correspond to a
higher magnetization and consequently to a larger gain in the
magnetic energy. Finally, it is interesting to observe that by
computing the spin-dependent reflection and transmission
probabilities for the weak ferromagnet regime, the processes become
almost equivalent thus making more difficult to distinguish
between the spin-polarization modification at the interface for
the parallel and the antiparallel configurations. Then, the
mechanism of wave-function reconstruction gets not effective in
such regime.

In order to
understand the origin of the antiparallel bulk-interface ground
state in the phase diagram of the FM-N heterostructure for a
weak ferromagnet,
 it is useful to analyze the problem by
reducing the model to only two layers, i.e. the interface layer
and its first neighbor (respectively $i_{z}$=0 and $i_{z}=-1$ in the
notation of the heterostructure) for the two magnetic
configurations. Assuming a collinear configuration, we consider
the change of the energy spectrum due to the hybridization of the
bands with the same spin polarization (the non collinear case does
not change the results qualitatively). For the parallel
configuration the spin polarized energy bands get split by the
interlayer hopping and in the limit of $h=h_{\text{int}}$ they are
given by $E_{\sigma}^{P}=\epsilon(k)-\sigma(\,h*M+t_{\perp})$. The
expression for $E_{\sigma}^{P}$ is obtained by assuming the same
in-plane spectrum within the layer at $i_z=0$ and $-1$, and the
index $\sigma=1 (-1) $ labels the spin majority (minority)
electrons in the bulk FM. For the antiparallel configuration the
energy spectrum is split as a square root of the interlayer
hopping, i.e. we have
$E_{\sigma}^{AP}=\epsilon(k)-\sigma\sqrt{(h*M)^2+t_{\perp}^2}$. By
comparing the energy spectra for the two magnetic configurations
for the spin-majority channels, one can deduce that for small
exchange fields the energy gain is larger for the parallel
configuration if compared to the antiparallel one because of the
linear correction in the energy spectrum.

We conclude that for weak-intermediate ferromagnets the parallel
bulk-interface magnetic configuration is the most energetically
favored state because of the energy spectrum reconstruction. On
the other hand, in the regime of strong ferromagnetic exchange field, it
is the wave function reconstruction that primarily drives the
stability of the antiparallel configuration. We finally note that,
for the FM-N heterostructure, the invariance of the Hamiltonian
under spin rotation makes all the magnetic orientations degenerate
in energy. These conclusions have been drawn for a physical
situation with collinear magnetization. Similar arguments apply
also for the case of a noncollinar bulk-interface configuration.
\\

We focus now on discussing the reasons for having the spin-triplet
superconductor to favor an antiparallel bulk-interface magnetic
state with a planar spin orientation (i.e. perpendicular to the
$\vec{d}$-vector) at any given strength of the ferromagnetic
exchange field. We recall that this behavior can be deduced by the
observation of the profile for the pairing energy $E_{\text{op}}$
which is always minimized when the magnetization in the bulk and
at the interface lies in the plane perpendicular to the
$\vec{d}$-vector and are aligned with opposite spin orientations.
Since the $E_{\text{op}}$ energy is directly related to the
amplitude of the order parameter in the superconducting region,
it is relevant to
analyze the intensity and the character of the Andreev processes.
Is is well known that the Andreev reflection amplitude enters
plays a role in this context because it
is the probability of an electron (hole) injected from the
ferromagnet to be reflected as hole (electron), thus providing
the intensity for the creation (annihilation) of an electrons
(holes) pair in the superconductor. Then the Andreev
scattering amplitude sets the strength of the pair fluctuations in
the superconductor close to the interface and thus it is the main
effect in driving the modification of the pair amplitude. Based on
the nature of this process, we can argue that a large value of the
Andreev reflection amplitude at the FM-TSC interface
would then result into a significant
reduction of the pair amplitude and in turn of
the superconducting order parameter in the TSC region close to the interface.

Among all the Andreev processes, those related to the creation
(annihilation) of an electrons pair with a different phase are of
particular relevance because they interfere destructively when
they sum up in the formation of the pairing amplitude. From a
general point of view, such non-trivial signs in the Andreev
scattering can originate from the orbital dependence of the
superconducting order parameter or they can be due to
spin-dependent phase shifts for the mismatch in the magnetic
configuration between the superconductor and the ferromagnet. A
distinct and limiting case occurs in the presence
of bound states at the interface where the Andreev probability
is maximized for the correspondent energy of the incident 
electron (hole). The relation between the
Andreev processes and the pairing amplitude variation has been
shown to hold for a $d$-wave superconductor interfaced to the
vacuum where bound states occur at specific orientations of the
interface with respect to the $d$-wave nodal directions. There, a
maximal reduction of the order parameter is observed when the
orientation of the interface is such to have Andreev bound states
in the density of states\cite{Kashiwaya2000}.

As mentioned in the Section III, for the case of a spin-triplet
superconductor and for the examined layered geometry in this
paper, the chiral orbital symmetry does not lead to Andreev
scattering processes that involve sign changes in the pairing
potential along the particle-hole trajectories. Hence, we expect
that the modifications of the order parameter should be mainly
ascribed to the contributions of spin-dependent scattering. Taking
into account the results for the homogeneous
ferromagnet\cite{Brydon2009,Brydon2011,Gentile2012}, we are aware
that the pair fluctuations are suppressed when the magnetization
of the ferromagnet lies in the plane of the spin of the Cooper
pairs, i.e. $E_{\text{op}}$ is maximal for a magnetization which
is perpendicular to the $\vec{d}$-vector. This is because in such
a configuration there are no phase shifts in the Andreev
processes\cite{Brydon2009}. 
Considering now the competition between the other possible
inequivalent configurations, i.e. the parallel and the
antiparallel bulk-interface magnetic states, we observe that, in general and in the
absence of bound states, the amplitude of the Andreev scattering
decrease as a function of the barrier potential
strength\cite{BTK}.
 We note again that in the antiparallel
configuration the incident electron in the spin-majority band gets
experience of a greater barrier potential if compared to case of
parallel spin-polarization. Such observation would imply that the
intensity of the Andreev scattering probability is smaller for the
antiparallel bulk-interface magnetic state than for the parallel
one \and therefore that the
pairing amplitude is less affected by the scattering processes and
thus less reduced. This qualitative argument would
support the results of having a larger gain in
the condensation energy when the bulk-interface magnetizations are
antiparallel oriented. We also note that the discussed mechanism
is consistent with the fact that there is a weak dependence on the
strength of the ferromagnetic exchange field.

\section{Conclusions}

We have studied the physical behavior of a ferromagnet-
spin-triplet superconductor heterostructure with a layered
geometry assuming an orbital chiral symmetry for the
superconducting order parameter and spin dependent scattering at
the interface. By means of a self-consistent approach on a
three-dimensional lattice we have determined the spatial profile
of the pairing amplitude and magnetization in terms of the
strength of the ferromagnetic exchange field and the spin-active
processes at the interface. We find that the pairing amplitude in
the TSC region close to the interface increases if the
magnetization moves from being parallel to perpendicular to the
$\vec{d}$-vector. In the presence of a spin-active interface a
configuration with in-plane antialigned magnetic moments for the
bulk-interface layers maximizes the value of the pairing
amplitude. As expected, the penetration of the Cooper pairs in the
FM leads to induced pairing amplitudes with zero-spin projection
that we show to have a damped oscillating behavior if the
magnetization is out-of-plane, while it is monotonously decaying
when the magnetization is perpendicular to the $\vec{d}$-vector,
i.e. parallel to the spin direction of the Cooper pairs. A
distinctive feature is the oscillating behavior of the proximity
effect which brings a phase change only if the angle between the
magnetization and the $\vec{d}$-vector is in the $\sim[0,\pi/4]$
window. The spin-active scattering mainly modifies the behavior
close to the interface by influencing the intensity and the phase
change of the pairing amplitude. One of our main
results has been to show that
spin-polarized pair correlations are also induced both at the
interface in the TSC and with a long range penetration length in
the FM if the magnetization is noncollinear to the
$\vec{d}$-vector. We have shown that their amplitude is maximized
at an angle $\theta\sim\pi/4$ for a uniform ferromagnet and it has
a non trivial behavior in the presence of a bulk-interface
magnetic misalignment. We have pointed out that such correlations
can play a major role in the case of the existence of a pair
potential in the same spin channel at the interface of the
spin-triplet superconductor. This effect can lead to a
re-orientation of the $\vec{d}$-vector approaching the ferromagnet
and drastically influence the anisotropy of the ferromagnet as
well as the charge and spin transport of the hybrid
structure\cite{Brydon2011}.\\

The knowledge of the key physical quantities of the FM-N and
FM-TSC hybrids has been used to provide the basis for a detailed analysis of the
optimal magnetic configuration as due to the coupling with a
spin-triplet superconductor. The study of the Gibbs free energy of
the system reveals that the rotational spin invariance of the
ferromagnet is generally broken due to the coupling with the
superconductor and in turn the FM develops an anisotropy which
favors an in-plane orientation of the magnetization. In the case
of a uniform ferromagnet, we have demonstrated that the driving
mechanism for the stabilization of the bulk-interface relative
orientation of magnetization is based on the gain of the
superconducting condensation energy rather than on the proximity
effect. The presence of an active ferromagnetic layer at the
interface leads to a different physical scenario. There are two
optimal magnetic states which minimize the Gibbs free energy of
the system: they have a parallel or antiparallel bulk-interface
alignment with spin orientation degeneracy in the FM-N hybrid
while with a magnetization oriented in the plane perpendicular to
the $\vec{d}$-vector for the FM-TSC heterostructure.
As our central result we have
demonstrated, by exploring
the full space of parameters as a function of the relative
exchange field strength at the interface and in the bulk,
that the system can switch between the two
ground-state configurations. The antiparallel configuration is
generally favored for a regime of weak and strong ferromagnet
otherwise the parallel state is stabilized. In the latter the spin
dependence of the energy spectrum in the FM domain close to the
interface cooperates with the energy contribution due to the
modification of the superconducting order parameter which would
always tend to favor an antiparallel interface-bulk magnetic
configuration. We emphasize that such electronic structure effect
has important contributions inside and outside the energy window
of the superconducting gap. We have demonstrated that such effects
lead to noncollinear bulk-interface magnetic configurations when
the bulk magnetization is assumed fixed along a given direction
and that at the interface can vary to minimize the Gibbs energy.\\

The results obtained in this paper represent a solid platform to
investigate a FM-TSC hybrid in a conventional epitaxially grown
geometry. We point out that they can be of direct interest for the
case of the Sr$_2$RuO$_4$ spin-triplet superconductor interfaced
with other ferromagnetic oxides having matching lattice
parameters, i.e. SrRuO$_3$ or doped LaMnO$_3$, but can also be
indirectly used to extract information of the symmetry of the
order parameter of unknown superconductors through the
investigation of the magnetic profile in the ferromagnet. Indeed,
the occurrence of magnetization anisotropy within the FM can be
experimentally probed measuring the magnetic moment profile as a
function of the layer position via nuclear magnetic resonance,
muon spin rotation or polarized neutron reflectometry. The latter
has been successfully applied, for instance, to observe a
surprisingly large superconductivity-induced modulation of the
ferromagnetic magnetization profile in superlattices made of high
temperature superconductors interfaced with ferromagnetic
manganites\cite{Hoppler09}. Finally, since a spin-active interface
plays a relevant role in determining the character of the
spin-triplet pair correlations induced in the FM and in tuning the
free energy profile, it is expected to be an important ingredient
for possible spin-valve devices based on spin-triplet
superconductors as well as for controlling the dynamical evolution
of the magnetization in the ferromagnet.

\section*{Acknowledgment}
The research leading to these results has received funding from
the EU -FP7/2007-2013 under grant agreement N. 264098 - MAMA. M.C.
and D.M. acknowledge support from the Short Term Mobility (STM)
program of the Consiglio Nazionale delle Ricerche (CNR).

\end{document}